# *Perfect Splitting in Rectangular Waveguide Junctions for Analogue Computing*

William Rogers[1,#], Christian Johnson-Richards[1, #] Alex Yakovlev [2]
and Victor Pacheco-Peña [1,*]

[1] School of Mathematics, Statistics and Physics, Newcastle University, Newcastle Upon Tyne, NE1 7RU, United Kingdom
[2] School of Engineering, Newcastle University, Newcastle Upon Tyne, NE1 7RU, United Kingdom

*email: victor.pacheco-pena@newcastle.ac.uk

**It has been recently shown how computing operations such as high-speed switching, routing, and solving partial differential equations can be performed by exploiting perfect splitting of electromagnetic waves in networks of waveguides from microwaves to the optical regime. Here, we propose a technique to achieve perfect splitting of electromagnetic waves using junctions of rectangular waveguides. The structure consists of *N* air-filled rectangular waveguides interconnected at a junction. They are designed to have their cut-off frequency above the cut-off frequency of further *N* waveguides used as inputs. The proposed structure is studied theoretically using transmission line models demonstrating that perfect splitting can occur at frequencies below the cut-off frequency of the interconnected waveguides (evanescent coupling). Numerical results are implemented to validate the designs demonstrating a good agreement between them. As examples of computing operations, it is shown how the proposed structure can be used to *compare* the amplitude of two incident signals (comparison operation) and also for routing of information to different output ports by exploiting linear superposition of scattered waves excited at the junction of rectangular waveguides, opening new opportunities and possibilities for future exploration and exploitation of electromagnetic waves for high-speed computing.**

---

[#] These authors contributed equally to this work.

# I. Introduction

The increasing need of powerful and faster computing systems has pushed the field of computing to evolve at an unprecedented speed. However, challenges in the performance and manufacturing of integrated circuit chips (e.g. chips are approaching physical size limits [1], [2]) need to be tackled to fulfill demands [3]. Classical computing systems, such as digital electronics, rely on a large number of transistors as switching devices. However, inherent to such devices, are unavoidable parasitic capacitances that appear at their gate. The charging and discharging of these capacitances during switching introduce delays and can cause energy dissipation, reducing the overall performance of computing system [4]. While there is ongoing research to address these challenges, there is a growing interest in the development of alternative computing paradigms, giving rise to interesting examples such as quantum [5], biological [6] and photonic computing approaches [7]–[15].

Within the realm of photonic computing, recent examples in this field include methods for solving quadratic optimization problems [7], differentiation and integration [8]–[12], optical neural networks [13], [16], [17] and partial differential equation solving with interconnected waveguide junctions [18], [19], among others, demonstrating how this paradigm may offer opportunities for low power, high-speed digital and analogue computations [20]–[29]. In this context, we have recently shown how ideal splitting of electromagnetic (EM) signals between a junction of 4-port parallel plate waveguides can be utilized for elementary computing operations such as information routing and comparison operations [30], [31], which provides a fundamental pathway for elementary decision-making computations. A similar approach has also been demonstrated using plasmonic junctions at optical frequencies [32]. Perfect splitting in a 4-port waveguide junction, as an example, requires an incoming signal applied from one of the waveguides to be split into four signals after passing the junction. Each signal (traveling within each of the waveguides) will have the same magnitude with a phase difference of $\pi$ between the reflected (traveling towards the incident port) and transmitted signals (traveling towards the other three ports) [33] (full details about such perfect splitting in *N*-port waveguide networks will be shown in the following sections). Such waveguide junctions require a large incident wavelength to waveguide width ratio in order to consider them as perfect (ideal) transmission lines (TLs), and this is the reason why they have been implemented mainly using parallel plate waveguides [30], [31], [34], [35]. Here, we build upon these works to enable elementary computing operations by implementing arrays of rectangular waveguides as realistic and technologically relevant structures. As in the ideal configurations, our work also relies on perfect splitting of EM waves at a waveguide junction. However, the large wavelength to waveguide width ratio, required to mimic ideal TLs for perfect splitting [26], is limited when working with rectangular waveguides due to their width (along the direction perpendicular to the *E*-field) is linked to the cut-off wavelength of the fundamental $TE_{10}$ mode and small cross-sections cannot be achieved [33]. However, this could be addressed by implementing alternative waveguide junctions using rectangular waveguides of carefully designed cross-sections.

Inspired by the importance of photonic computing and the need for perfect splitting in arrays of interconnected waveguides, here we present our efforts to achieve perfect splitting of incident EM signals within a 4-port network of rectangular waveguides operating below their fundamental $TE_{10}$ mode. The structure consists of four vacuum-filled rectangular waveguides designed to have their cut-off frequency at $f_{c_w}$ (called outer waveguides). These waveguides are interconnected via a cross-junction comprised of four vacuum-filled rectangular waveguides having a higher cut-off frequency, $f_{c_x} > f_{c_w}$ (called junction waveguides). In so doing, it will be shown how perfect splitting of EM waves can be achieved at frequencies that fall above the cut-off frequency of the outer waveguides but below the cut-off frequency of the junction waveguides (i.e. evanescent coupling). The mechanisms enabling such perfect splitting is presented and studied via numerical simulations using CST Studio Suite®. To understand this performance, in-depth analytical studies using TL theory are presented. It is shown how, the reactive field at the junction can be modelled via a reactive element demonstrating how there are frequencies where perfect splitting is achieved even below the cut-off frequency of the junction waveguides, in agreement with numerical simulations. For completeness, we show how the proposed configuration using rectangular waveguides can be exploited for *N*-port perfect splitting. Finally, we implement the proposed structure to perform some computing operations including elementary signal routing and amplitude comparison operations.

## II. Principles of perfect splitting and proposed structure

### A. Definition of perfect splitting.

As mentioned in the introduction, perfect splitting of EM signals has recently been exploited for fundamental computing operations such as routing of information [30], [31], [36]. A detailed explanation of the implications of perfect splitting using interconnected TLs can be found in [30], but a summary is provided here for completeness. The schematic representation of the concept of perfect splitter of EM signals considered in this work is presented in Fig. 1a, where it is shown how an incident signal applied from port 1 is split into *N* signals, each of them traveling towards each of the ports connected to the ideal splitter. Note that the concept of perfect splitting considered here is different than multiport power dividers where the incident signal is equally divided and delivered to the *N*-1 output ports of a network without reflection [37]. Here, however, a reflected signal travelling towards port 1 will also be necessary to be considered as perfect splitting. A perfect splitter can be designed by using ideal TLs interconnected at a junction either in series or parallel configuration [31], [36]. From TL line theory [33] if all the interconnected TLs have the same impedance, $Z_1 = Z_2 = Z_3 \ldots Z_N = Z$, the incident signal applied from port 1 (TL with impedance $Z_1$) will see a change of impedance from $Z_1 = Z$ to $Z_{out} = (N-1)Z$ or $Z_{out} = \left(\frac{N-1}{Z}\right)^{-1}$ if the TLs are interconnected in series or parallel configuration, respectively. In so doing, transmitted and reflected signals will be traveling towards the output and input ports, respectively with a transmission and reflection coefficient (for each port) defined as $\gamma = -2N^{-1}$ and $\rho = (N-2)N^{-1}$, respectively, if the TLs are interconnected in series, or $\gamma = 2N^{-1}$ and $\rho = (2-N)N^{-1}$, respectively, if they are connected in parallel (the latter case is depicted in Fig. 1a). Note that the signs of $\gamma$ and $\rho$ have been written to directly account for the change of phase of the transmitted and reflected signals [30]. Importantly, for the case when four TLs are interconnected in a parallel configuration (as it will be the case in the next section), the reflected signal has the same amplitude, but it is 180° ($\pi$) out of phase with the transmitted signals.

Now, if multiple signals are applied from different ports, one can simply exploit the linearity of the system and consider that the output signals towards each port will be the superposition of the scattered signals excited after each of the input EM signals have passed the junction. i.e., for an input vector of EM signals $\pmb{x} = [x_1, x_2, \dots x_N]$, with subscript indicating the port number, an output vector of EM signals $\pmb{y} = [y_1, y_2, \dots y_N]^T$ can be defined. This means that, for the case of $N$ TLs interconnected in series or parallel configuration, $\pmb{y}$ can be simply calculated by multiplying an $N \times N$ scattering matrix with diagonal and off-diagonal terms representing the reflection and transmission coefficient of the junction, respectively, and the input vector of signals, as:

$$\pmb{y} = \begin{pmatrix} \rho & \gamma & \gamma & \cdots \\ \gamma & \rho & \gamma & \cdots \\ \gamma & \gamma & \rho & \gamma \\ \vdots & \vdots & \gamma & \ddots \end{pmatrix} \pmb{x}^T \qquad (1)$$

with $T$ as the transpose operator. To implement the perfect splitting theory described above and to exploit it for computing operations, one can make use of waveguides as TLs [33]. However, for them to behave as ideal TLs (without exciting fringing fields at the junction), it is required to have a much larger incident wavelength of the EM signal compared to the cross-section of the waveguide. It has been shown that when this ratio exceeds ~8:1, interconnected waveguides can mimic ideal interconnected TLs [34]. This is possible with, for instance, parallel plate waveguides where the distance between the metallic plates can be reduced to fulfill this condition. However, parallel plate waveguides interconnected at a junction are ideal structures that require, for instance when connected in series, infinitely long plates along the direction orthogonal to the propagation of the EM signals. This is to avoid fringing fields at the edges of the waveguides and at the junction (i.e., ideal 2-dimensional structures). One may ask, is there a way to emulate perfect splitting using rectangular waveguides instead, as examples of technologically relevant structures? At first glance, one can envision that this can become a challenge provided that the dimensions of rectangular waveguides are of the order of the cut-off wavelength, $\lambda_c$, of the fundamental $TE_{10}$ mode. In fact, TE modes only exist when the incident signal wavelength, $\lambda$, is smaller than $\lambda_c$. The cut-off wavelength (and hence frequency) of the first $TE_{10}$ mode, $\lambda_c$, is defined by the width, $a$, of the waveguide (dimension transversal to the direction of the electric $E$-field), as $\lambda_c = 2a$ [35]. Subsequently, the cut-off wavelength limits the maximum wavelength to waveguide width ratio to 2:1 when working near cut-off. To demonstrate this, an example showing four interconnected waveguides at a junction is shown in the supplementary materials, demonstrating how there are no frequencies above the cut-off frequency of the fundamental mode where perfect splitting can be achieved, as expected. Based on this, is there a possibility to overcome this using rectangular waveguides? In the following section we address this question by proposing an alternative configuration for interconnected waveguides.

**B. Perfect splitting: interconnected rectangular waveguides.**

The proposed structure to enable perfect splitting of EM signals using interconnected rectangular waveguides is shown in Fig. 1b. Here we first focus on a 4-port configuration, further examples using $N$ interconnected waveguides will be presented in the following sections. As observed in Fig. 1b, the structure is composed of four outer waveguides, interconnected by four waveguides forming a junction (junction waveguides as defined in the introduction). All the rectangular waveguides (outer and junction waveguides) are filled with vacuum ($\mu_r = \varepsilon_r = 1$) and surrounded by perfect electric conductor (PEC). The width of the waveguides are chosen following the same dimensions as in [38] to have the cut-off frequencies for the outer waveguides and the junction waveguides as $f_{c_w} \sim 1.043$ GHz and $f_{c_x} \sim 1.475$ GHz, respectively, with $f_{c_w} < f_{c_x}$. The widths of the outer and junction waveguides (direction perpendicular to the *E*-field) are $a_{w,x} = \frac{\lambda_{c_{w,x}}}{2}$, respectively, where $w$ and $x$ refer to the outer waveguides and the junction waveguides respectively, and $\lambda_c$ is the cut-off wavelength of the respective waveguide. The heights (in the direction parallel to the *E*-field), are $b_w = \frac{a_w}{2}$ and $b_x = a_x \Delta h$, initially with $\Delta h = 1/64$, again in line with [38]. The length of the outer waveguides is $\ell_w = 2a_w$ (an arbitrary value as waveguide ports will perfectly absorb the incoming waves) while the length of junction waveguides is chosen to be $\ell_x = 1.2303 \lambda_{c_x}$. It is important to note that this is the total length of the junction waveguides measured from outer waveguide to outer waveguide (see Fig. 1b).

With this configuration, the proposed structure in Fig. 1b was numerically evaluated using the frequency domain solver from the commercial software CST Studio Suite®. To do this, four rectangular waveguide ports were added to the outer waveguides. A tetrahedral mesh with adaptive refinement was implemented with a maximum cell density of 13 cells per $\lambda$ inside and outside the model. Electric boundary conditions were set on all model boundaries to consider PEC. For the regions that cannot touch the boundaries (e.g., around the junction waveguides as their dimensions are smaller than the outer waveguides) PEC was used as the background medium to ensure that all the waveguides are covered with PEC. Finally, the frequency domain solver was configured with a third order accuracy and the scattering parameters were calculated for a maximum number of 100001 frequency samples within the frequency range $\sim 0.67 \leq f/f_{c_x} \leq \sim 1.35$.

The numerical results of the scattering parameters for the proposed 4-port structure are presented in Fig. 1c(i-iv) for ports 1 – 4, respectively. The scattering spectra have been normalized with respect to the cut-off frequency of the junction waveguides $f_{c_x}$, and the results are plotted for the normalized frequencies $0.7 \leq f/f_{c_x} \leq 1.3$ to account for incident frequencies above and below the cut-off frequency of the fundamental (TE$_{10}$) and higher order mode (TE$_{20}$), respectively, of the outer waveguides. As observed, there are clear resonant frequencies where the amplitude at each port is ~0.5 (at normalized frequencies $f/f_{c_x}$~0.7967, ~1.1155, and ~1.2404), which is a condition for perfect splitting, as described in the previous section and in Eq. 1. Interestingly, there is a resonant frequency ($f/f_{c_x}$ = 1.0626) where most of the incident signal is transmitted towards port 3, i.e., ~1 transmission straight through. This latter resonance, while not used here for perfect splitting, may be attributed to the junction waveguides working as effective media emulating epsilon-near-zero metamaterials [38]–[43]. To further study the resonances shown in Fig. 1c(i-iv), the out-of-plane $E_y$-field distributions were extracted, and the results are shown in Fig. 1c(v-viii) [to guide the eye, see also labels v-viii in Fig. 1c(ii) for each resonant frequency]. As observed, the field distribution inside the junction waveguides resembles a line source along the out-of-plane direction for the first resonance [Fig. 1c(v)]. For the resonant frequencies where the amplitude of the transmitted/reflected signals is ~0.5 [Fig. 1c(v,vii,viii)], the phase difference between the transmitted signal and the reflected signals is ~π for the first and fourth resonance [Fig. 1c(v,viii), respectively] with values of $\Delta\phi_{21}$~0.999π and ~1.006π, respectively, when calculating it using port 2 and port 1. However, this is not the case for the third resonance [Fig. 1c(vii)] where a phase different of ~π is only achieved between port 2 and the rest of the ports (i.e., non-ideal splitting). See supplementary materials for more details about the phase at each port for each of the resonances from Fig.1. In this work, while the resonance appearing at a higher frequency [fourth resonance, Fig. 1c(viii)] also enables perfect splitting, we focus our attention on the first resonance [Fig. 1c(v)] as it appears at a lower frequency, enabling the implementation of a more compact computing structure relative to the operating wavelength. Notably, the first resonance which enables perfect splitting of EM signals, appears at a frequency $f_{c_w} < f < f_{c_x}$, (i.e., below the cut-off frequency of the junction waveguides) meaning that this resonance is the result of the coupling of an evanescent field to the junction waveguides. In the next section, we provide a theoretical analysis of the structure shown in Fig. 1b based on TL theory to understand how the reactive evanescent field can be modelled using equivalent circuit models.

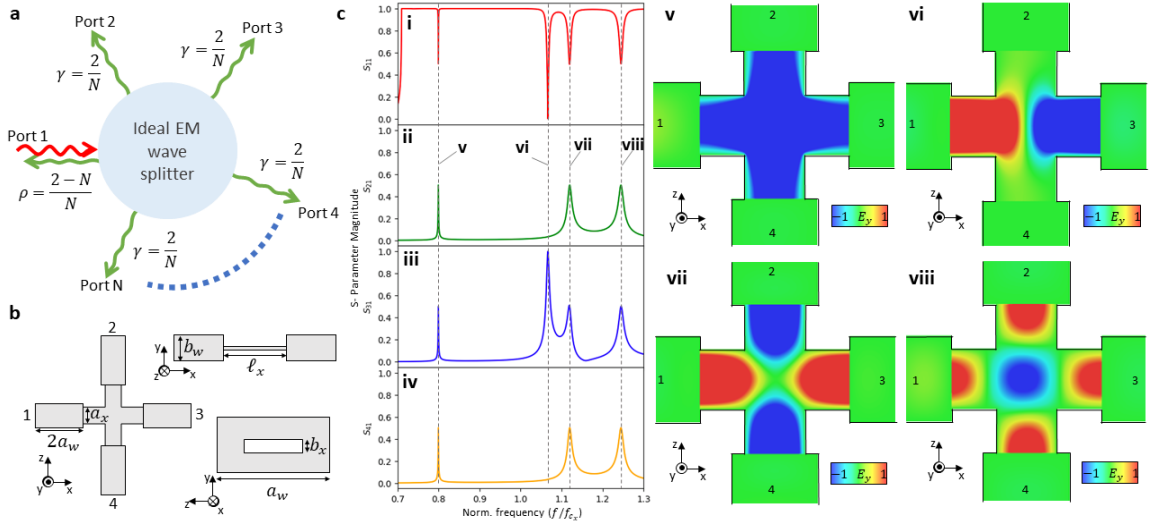

**Fig. 1 Proposed idea for perfect splitting using interconnected rectangular waveguides:** (a) Schematic representation of an ideal splitter in a *N*-port device. (b) Proposed structure for perfect splitting of EM waves. The structure consists of outer waveguides with a cut-off frequency of $f_{c_w} \sim 1.043$ GHz (with height and width $b_w = \frac{a_w}{2}$ and $a_w = 0.1016\sqrt{2}$, respectively) connected to a junction of waveguides having a cut-off frequency of $f_{c_x} \sim 1.475$ GHz (with height, width and length $b_x = a_w \Delta h$, $a_x = \frac{a_w}{\sqrt{2}}$, and $\ell_x \sim 1.2303 \lambda_{c_x}$). (c) (i-iv) Scattering-Parameters for port 1, 2, 3 and 4, respectively, considering an input signal from port 1 calculated using CST Studio Suite®. (c) (v-viii) $E_y$-field distribution on the *xz* plane (*H*-plane) at each of the resonances shown in panels i-iv (see labels in panel ii).

### III. Theoretical analysis

In the previous section, it has been shown how perfect splitting of EM signals can be achieved using interconnected rectangular waveguides as proposed in Fig. 1b. This section is focused on the theoretical analysis of the proposed structure using TL theory, and it is divided as follows: first, in subsection A, an equivalent 2-port TL model for the structure shown in Fig. 1b is considered, where three of the four *arms* (here an *arm* refers to an outer waveguide + a junction waveguide connected at the junction, i.e., Fig. 1b has a total of four *arms*) are grouped as an equivalent impedance. In this model, as it will be shown, an interface admittance and a junction admittance are considered, to model the reactive fields at the interface between an outer and a junction waveguide and the reactive field at the junction between waveguides, respectively. The effect of these parallel admittances will be discussed in detail. Next, in subsection B, we build upon the equivalent 2-port circuit from subsection A and present a technique to calculate the parallel admittance modelling the reactive field at the junction waveguides. It will be shown how such admittance can be calculated using a combination of numerical results (simulation) and analytical calculations via TL models and the ABCD matrix method. Finally, in section C, it is shown how the proposed structure from Fig. 1b can be also exploited for perfect splitting using *N*=3, 5, 6 interconnected waveguides.

**A. Effect of parallel interface and junction admittances: equivalent circuit model.**

The equivalent circuit for the proposed structure shown in Fig. 1b is presented in Fig. 2c. Each waveguide, outer and junction waveguides (all filed with air $\mu_r = \varepsilon_r = 1$), are modeled as TLs having impedance, propagation constant, and length $Z_{w,x} = \frac{\omega}{\beta_{w,x}} \frac{b_{w,x}}{a_{w,x}}$, $\beta_{w,x} = k_0 \sqrt{1 - \left(\frac{\pi}{k_0 a_{w,x}}\right)^2}$, $\ell_w$ and $\ell_x/2$, respectively, with $\omega$ as the angular frequency, and the free space wave number $k_0 = 2\pi f \sqrt{\mu_0 \varepsilon_0}$, where $f$ is incident frequency, $\varepsilon_0$ is the permittivity and $\mu_0$ the permeability of free space respectively. Note that in the TL model the length of each junction waveguide is considered to be $\ell_x/2$ to account for their length from the outer waveguides to the center of the junction ($\ell_x$ is measured from outer waveguide to outer waveguide, as described in Fig. 1b). Finally, in this manuscript the time convention $e^{-i\omega t}$ is used. As mentioned in the previous section, $\ell_w$ can have an arbitrary value as the outer waveguides are finished with absorbing ports in the numerical analysis, hence, in the TL model they are simple impedances connected to the circuit [33]. The structure shown in Fig. 1b has two discontinuities where reactive fields can be considered [38]: i) one at the interface between each of the outer waveguides and the junction waveguides [where dimensions change from $(a_w, b_w)$ to $(a_x, b_x)$ along the $(x,y)$ axes, respectively], and ii) another at the center of the junction waveguides. To model the reactive fields at the interface between the outer and junction waveguides, one can place a parallel admittance between the two TLs modelling these waveguides. This is represented in Fig. 2c where a parallel admittance $Y_I = -i\Psi_I$ (interface admittance, with $\Psi_I$ representing the coefficient of the admittance) is placed between $Z_w$ and $Z_x$ for each of the four *arms* in the TL model. The nature of $Y_I$ can be explained by considering the theory of waveguide discontinuities where a change of width and height at the interface between the waveguides can be modelled as an equivalent LC parallel circuit [38]. Similarly, a parallel admittance $Y_J = -i\Psi_J$ (junction admittance, with $\Psi_J$ as the coefficient of the admittance) is also placed at the junction of waveguides connecting the four *arms* to account for the reactive fields at the center of the junction, noting that such admittance could represent, for instance, a parallel capacitance [33], [44].

To analyze the equivalent circuit from Fig. 2c, for simplicity, the three output *arms* and $Y_J$ from Fig. 2c (right panel) can be combined into a single impedance, $Z_{tot}$, (see schematic representation on the bottom panel from Fig. 2c), as follows [33]:

$$Z_{tot_1} = \left(Y_J + \frac{3}{Z_{in}}\right)^{-1} \tag{2}$$

where $Z_{in}$ is the input impedance seen from the center of the junction towards an output port. From TL theory, $Z_{in}$ can be expressed as [44]:

$$Z_{in} = Z_x \frac{Z_{par_1} - iZ_x \cos(\beta_x \ell_x/2)}{Z_x - iZ_{par_1} \cos(\beta_x \ell_x/2)} \tag{3}$$

with $Z_{par_1} = \left(Y_I + \frac{1}{Z_w}\right)^{-1}$. With these simplifications, the resulting circuit model is reduced to (see bottom panel from Fig. 2c) an input *arm* (consisting of a TL modelling the single outer waveguide connected to port 1, a parallel admittance representing the interface between the outer and junction waveguide $Y_I$, and a TL representing the junction waveguide for this arm) connected to the load impedance, $Z_{tot_1}$. To extract the scattering parameters of this simplified 2-port TL circuit, the ABCD matrix method is implemented [33]. Based on this, the TL representing the junction waveguide of the input *arm* and the interface admittance $Y_I$ from the bottom panel of Fig. 2c can be defined by the following 2 × 2 matrices:

$$ABCD_x = \begin{bmatrix} \cos(\beta_x l_x/2) & -iZ_x \sin(\beta_x l_x/2) \\ -i\frac{1}{Z_x}\sin(\beta_x l_x/2) & \cos(\beta_x l_x/2) \end{bmatrix} \quad (4a)$$

$$ABCD_{Y_I} = \begin{bmatrix} 1 & 0 \\ Y_I & 1 \end{bmatrix} \quad (4b)$$

respectively [33]. An overall total ABCD matrix defining the circuit from Fig. 2c (bottom panel) can then be obtained by simply multiplying the two matrices described above. i.e., $ABCD_{eq} = ABCD_{Y_I} \cdot ABCD_x$. The scattering parameters for the reflection and transmission coefficients, $S_{11}$ and $S_{21}$, for the 2-port equivalent circuit from Fig. 2c (bottom panel) can then be calculated as [45]:

$$S_{11} = \frac{A_{eq}Z_{tot_1} + B_{eq} - C_{eq}\overline{Z_w}Z_{tot_1} - D_{eq}\overline{Z_w}}{A_{eq}Z_{tot_1} + B_{eq} + C_{eq}Z_w Z_{tot_1} + D_{eq}Z_w} \quad (5a)$$

$$S_{21} = \frac{2\sqrt{\text{Re}\{Z_w\}\text{Re}\{Z_{tot_1}\}}}{A_{eq}Z_{tot_1} + B_{eq} + C_{eq}Z_w Z_{tot_1} + D_{eq}Z_w} \quad (5b)$$

where $\overline{Z_w}$ denotes the complex conjugate operator, $A_{eq}, B_{eq}, C_{eq}$ and $D_{eq}$ are the components of the $ABCD_{eq}$ matrix, and Re{ } denotes the real component [45].

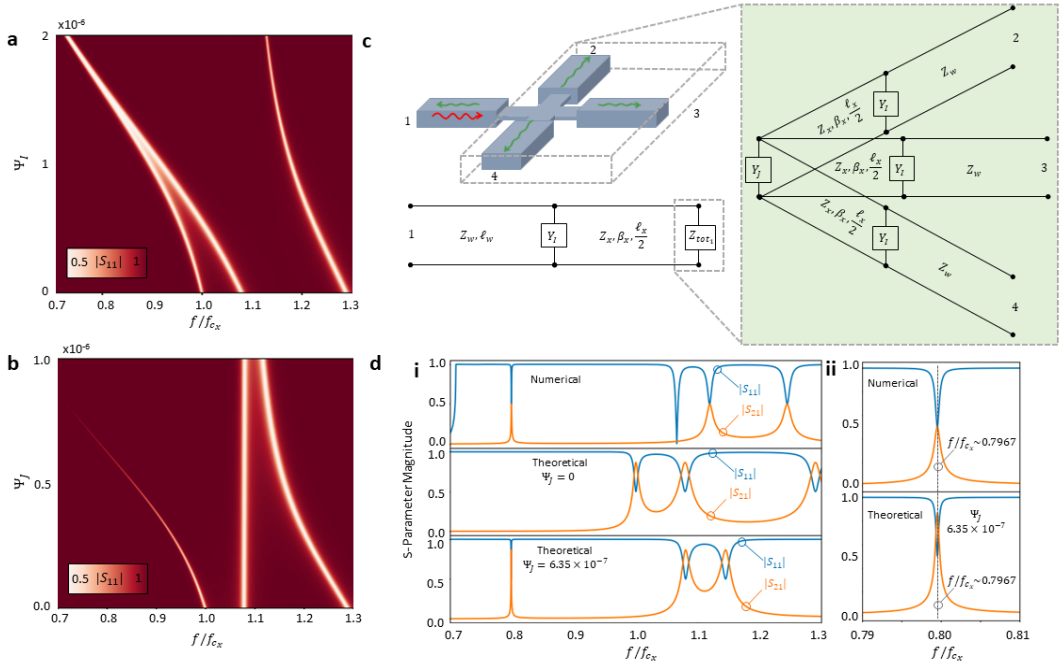

**Fig. 2| TL line model of the proposed 4-port waveguide junction for perfect splitting: lumped element representation of output *arms*.** (a) Theoretical results of $|S_{11}|$ as a function of the coefficient $\Psi_I$ of the interface admittance, $Y_I = -i\Psi_I$, and normalized frequency, considering $\Psi_J = Y_J = 0$. (b) Theoretical results of $|S_{11}|$ as a function of the coefficient $\Psi_J$ of the junction admittance, $Y_J = -i\Psi_J$, and normalized frequency, considering $\Psi_I = Y_I = 0$. (a,b) are calculated using the circuit from (c). (c) Schematic representation of a 4-port perfect splitter (same structure as Fig. 1b) and an equivalent TL model, consisting of three *arms* (highlighted in the green box) with parameters of outer waveguide impedance $Z_w$, outer waveguide length $\ell_w$, interface admittance $Y_I$, junction waveguide impedance $Z_x$, propagation constant $\beta_x$ and junction waveguide length $\ell_x/2$. The arms are connected to a parallel admittance $Y_J$. (d) i) Scattering-parameters calculated (top) using CST Studio Suite® for an input signal at port 1 and output to port 2, (middle) using the TL model with $\Psi_J = \Psi_I = 0$ and (bottom) $\Psi_I = 0, \Psi_J = 6.35 \times 10^{-7}$. For the theoretical calculations the scattering parameters represent the case using an input signal from port 1 and a combined output modelled by merging three *arms* (three output ports combined). ii) zoom-in results from the top and bottom panels from i) around the first resonant frequency.

Once a circuit model has been defined, the main challenges are now to calculate $Y_I$ and $Y_J$ and to understand how they affect the overall response of the equivalent circuit. To tackle this, a study into the influence of both $Y_I$ and $Y_J$ upon the scattering parameters of the TL model from Fig. 2c (bottom panel) was carried out via a parametric study of their corresponding coefficients ($\Psi_I$ and $\Psi_J$), respectively. First, a value of $\Psi_J = 0$ (i.e., $Y_J = 0$) was considered while $\Psi_I$ (and hence $Y_I$) was modified. The theoretical results of the magnitude of the $S_{11}$ as a function of normalized frequency and $\Psi_I$ are displayed in Fig. 2a. As observed, when $\Psi_I = \Psi_J = 0$ ($Y_I = Y_J = 0$) there are three resonant frequencies where $|S_{11}|{\sim}0.5$ (normalized frequencies $f/f_{c_x}{\sim}1.0000$, $\sim 1.0791$, and $\sim 1.2895$), i.e., all the resonances appear above the cut-off frequency of the junction waveguides $f_{c_x}$. As $\Psi_I$ is increased, the resonances are red shifted below $f_{c_x}$ with the two lower resonances merging for larger values of $\Psi_I$. Note that this red shift is a positive outcome as it was shown in the previous section how perfect splitting should appear for frequencies, $f_{c_w} < f < f_{c_x}$. However, manipulating $\Psi_I$ does not produce a spectrum that resembles the numerical results as the first two peaks merge when increasing this parameter.

A second study was carried out considering that $\Psi_I = 0$ (i.e., $Y_I = 0$) while varying $\Psi_J$ (and hence $Y_J$). The theoretical results of the $|S_{11}|$ as a function of normalized frequency and $\Psi_J$ are shown in Fig. 2b. As observed, as $\Psi_J$ is increased, the lower (first) and higher (third) frequency resonances are also red shifted (similar to the results shown in Fig. 2a) while the second resonance stays approximately at the same normalized frequency. For completeness, the numerical results of the $|S_{11}|$ (extracted from Fig. 2b) and $|S_{21}|$ considering $\Psi_J = 0$ (referred as case I) and $\Psi_J = 6.35 \times 10^{-7}$ (referred as case II) are shown in the middle and bottom panels from Fig. 2d(i), respectively, for each case along with the numerical results using CST Studio Suite® on the top panel. Note that the value of $\Psi_J$ for case II was manually chosen as a "guess solution" considering the best fit to the spectral location of the lower resonance from the numerical results. Zoom-in panels of the scattering parameters for the lower resonant frequency considering the numerical and theoretical results for case II are also shown in Fig. 2d(ii) (top and bottom panels, respectively) to guide the eye. As observed, the resonance appears at approximately the same spectral location ($f/f_{c_x} \sim 0.7967$) with values of $|S_{11}| \sim 0.5$ and $|S_{21}| = 0.86 \sim \sqrt{0.75}$. These results are as expected as the theoretically calculated $|S_{21}|$ considers three *arms* as a single output impedance (contained within Eq. 2). Hence, the transmission coefficient towards each of the ports (2 to 4) for each *arm* will then be $\left(\frac{|S_{21}|^2}{3}\right)^{1/2} = 0.5$, as the signal will be equally divided into the three *arms*. Importantly, apart from the location of the lower resonant frequency (first resonance), the other resonances do not agree with the numerical calculations [see Fig. 2d(i) top and bottom panels)] meaning that a frequency dependent $\Psi_J(f/f_{c_x})$ [i.e., $Y_J(f/f_{c_x})$] is necessary to model the full spectral response of the proposed structure. This will be explored below.

**B. Retrieval of parallel admittances: solving transmission line equations**

With a TL model in good agreement with numerical calculations at the frequency of the first resonance, we can now address the challenge of calculating a frequency dependent function for the admittance, $Y_J(f/f_{c_x})$, to match the full transmission spectra with the numerical results. In this section, and in the rest of the calculations, we consider $\Psi_I = 0$ as it has been shown above that the modelling of the perfect splitting performance can be achieved by simply using $\Psi_J \neq 0$, i.e., $Y_J = -i\Psi_J$.

With this in mind, Eq. 5a for $S_{11}$ [of the circuit model from Fig. 2c (bottom panel)] can be rearranged for $Y_J$, as follows: (see full derivation in the supplementary materials):

$$Y_J = \frac{A_{eq}S_{11} - A_{eq} + C_{eq}S_{11}Z_w + C_{eq}\overline{Z_w}}{B_{eq} - B_{eq}S_{11} - D_{eq}\overline{Z_w} - D_{eq}S_{11}Z_w} - \frac{3}{Z_{in}} \quad (6)$$

As observed, $Y_J$ is now dependent on $S_{11}$ and frequency, as required in this study. Now, as the input impedance $Z_{in}$ is complex valued, the resulting $Y_J$ is also complex. However, since the admittance represents a reactive element, the solution must be purely imaginary (i.e., a real $\Psi_J$).

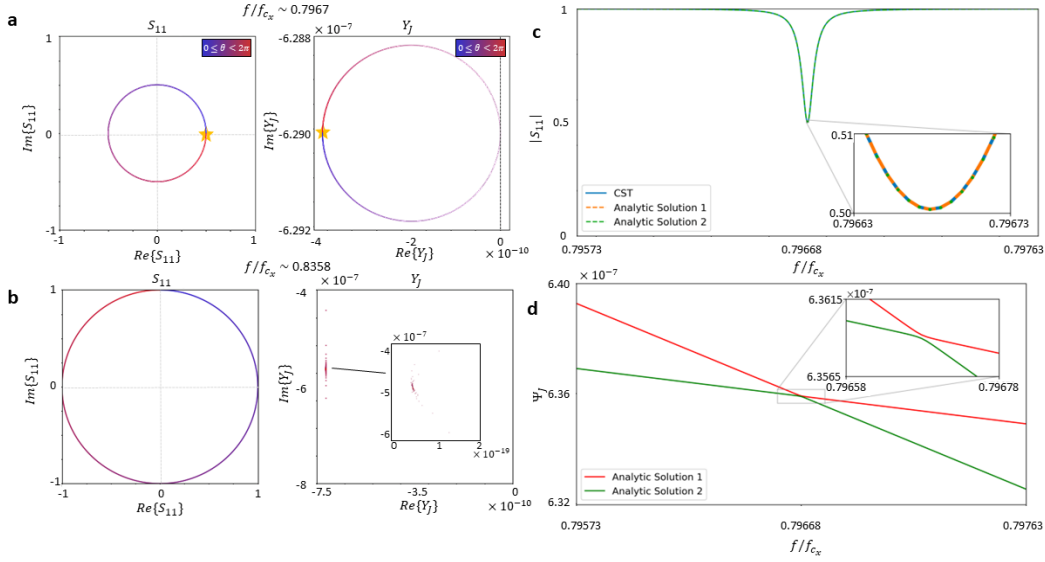

**Fig. 3| Calculation of purely imaginary $Y_J$: interception and conic section methods.** (a,b) Examples of the interception methods. The complex $S_{11}$ was numerical calculated and then rotated in the complex plane to obtain conic section of constant magnitude $|S_{11}|$. Results are shown on the left panels. Two frequencies are shown as representative examples: (a) at the frequency where perfect splitting of EM signals occurs $f/f_{c_x}\sim 0.7967$, and (b) $f/f_{c_x}\sim 0.8358$ (arbitrary frequency). The calculated complex $Y_J$ are shown in the right panels with the star representing the value of $Y_J$ calculated directly from the numerically extracted $S_{11}$ (before rotation). (c,d) Conic section method. (c) Numerical (blue) and theoretical (dashed lines) results of the $|S_{11}|$ around the first resonance where perfect splitting occurs. (d) Theoretical values of $\Psi_J$ using Eqs. 7-9 (red and green lines showing two solutions).

To better study the implications of Eq. 6, the value of $Y_J$ was extracted using the $S_{11}$ from the numerical simulations at the frequency where perfect splitting occurs (i.e., $f/f_c = 0.7967$). The resulting complex value is plotted in Fig. 3a (right panel) and plotted as a star symbol. For completeness the $S_{11}$ from the numerical simulation is also shown in the complex plane on the left panel of the same figure (again, shown as a star representing the extracted value from the numerical results). The resulting $Y_J$ is complex, as explained above. To overcome this, a method for calculating the equivalent real admittance was developed, as follows. For a given frequency, the complex $S_{11}$ was rotated in the complex plane by $2\pi$, generating 1000 values of $S_{11}$ with equal magnitude (see left panel from Fig. 3a for all the rotated values). For each of the values of $S_{11}$, the corresponding value of $Y_J$ was calculated using Eq. 6, generating values of $Y_J$ that form a conic section (see right panel from Fig. 3b, note that the $x$ and $y$ axis from this panel are different, meaning that the plot is indeed an ellipse). From these conic sections (one for each frequency), the interesting values of $Y_J$ are those that intercept, or satisfy, $Re\{Y_J\} = 0$ (black dashed line in Fig. 3a right panel), which can be found through interpolation. As a result, two purely imaginary valued admittances $Y_J$ can be obtained that will satisfy $|S_{11}|$ at a given frequency. However, this is not universal for all frequencies. As shown in Fig. 3b, there are frequencies (such as $(f/f_{c_x} = 0.8358)$ where the calculated values of $Y_J$ do not cross the imaginary axis (see right panel from Fig. 3b), preventing the use of this method of interception for the full spectrum. To find the intercept in all cases, however, one can use the full equation for a conic section [46]:

$$ax^2 + bxy + cy^2 + dx + ey + 1 = 0 \qquad (7)$$

where $a, b, c, d,$ and $e$ are constants to be found and $x$ and $y$ are the real and imaginary values of $Y_J$, respectively. To calculate these constants, five points from the generated admittances can be taken, $[(x_1, y_1), \dots, (x_5, y_5)]$, to form the following matrix equation:

$$\begin{bmatrix} x_1^2 & x_1y_1 & y_1^2 & x_1 & y_1 \\ x_2^2 & x_2y_2 & y_2^2 & x_2 & y_2 \\ x_3^2 & x_3y_3 & y_3^2 & x_3 & y_3 \\ x_4^2 & x_4y_4 & y_4^2 & x_4 & y_4 \\ x_5^2 & x_5y_5 & y_5^2 & x_5 & y_5 \end{bmatrix} \begin{bmatrix} a \\ b \\ c \\ d \\ e \end{bmatrix} = M \begin{bmatrix} a \\ b \\ c \\ d \\ e \end{bmatrix} = \begin{bmatrix} -1 \\ -1 \\ -1 \\ -1 \\ -1 \end{bmatrix} \qquad (8)$$

where $M$ is used to represent the matrix on the left-hand-side of this equation. To find the constants, the inverse of $M$ must be calculated, and then multiplied on the left to both sides of the equation, giving:

$$M^{-1}M \begin{bmatrix} a \\ b \\ c \\ d \\ e \end{bmatrix} = \begin{bmatrix} a \\ b \\ c \\ d \\ e \end{bmatrix} = M^{-1} \begin{bmatrix} -1 \\ -1 \\ -1 \\ -1 \\ -1 \end{bmatrix} \qquad (9)$$

Using the equation of the ellipse (Eq. 7) and the constants from Eq. 9, the intersection of the imaginary, $y$, axis can be found by setting $x = 0$, and solving the resulting quadratic equation $cy^2 + ey + 1 = 0$, leading to two solutions for a purely imaginary $Y_J$ that satisfies the numerically calculated $|S_{11}|$ at a given frequency. Based on this, the theoretically calculated values of $\Psi_J(f/f_{c_x})$, obtained from $Y_J$ using the method described above are shown in Fig. 3d for frequencies around the first resonance (where perfect splitting occurs). For completeness, from the two solutions of $Y_J$ (or its coefficients $\Psi_J$) (Fig. 3d), one can also calculate $|S_{11}|$ theoretically using Eq. 5a. The results are shown in Fig. 3c where it is clear how there is a perfect alignment with the numerical results (blue line). From these results, it will be expected an agreement between the numerical and theoretical values for $|S_{21}|$. As shown in the supplementary materials, the theoretically calculated $|S_{21}|$ using this technique and the numerically calculated spectra are in agreement with the total transmission at the junction for the first, third and fourth resonances. However, due to the unequal splitting nature of the second resonance (where ~1 straight through transmission is obtained, as discussed in Fig. 1), a different circuit model would be required to fit the transmission spectrum around this resonance. This could be done, for instance, by considering series impedances (one per *arm*) connected to the junction parallel admittance $Y_J$. In this context, each of these series impedances could be different to account for non-equal splitting between all the *arms*. However, this is out of the scope of the present work as our aim is to exploit the proposed structure for perfect splitting of EM signals. As discussed, this method of implementing a conic section enables us to calculate values of $Y_J$ (or its coefficient $\Psi_J$), that satisfy the numerically calculated scattering parameters. However, as this is solved using only the perfect splitting magnitude requirement, one cannot assume that the phase condition for perfect splitting will also be satisfied. In the supplementary materials, we explore an alternative technique to retrieve $Y_J$ for the first resonant frequency where the perfect splitting constraints are directly taken into account, such as the phase difference between output and input ports and voltages at each port.

## C. Perfect splitting: *N*-port configurations

The previous sections have been focused on the study of the structure from Fig. 1b using 4-port configuration of interconnected rectangular waveguides for perfect splitting. The aim in this subsection is to demonstrate how the proposed structure can also be used for an arbitrary number of ports *N*. To do this, one should consider the following: i) as discussed in Fig. 1-3, an incident signal applied from an input port will *see* a different impedance at the center of the junction waveguides, an impedance which will depend on the number of *N*-1 arms connected to the junction (see Fig. 2). Due to this, it would be expected that, if *N* is changed, the resonant frequency at which the perfect splitting occurs will also shift. ii) Moreover, the role of the dimensions of the cross-section and the length of the outer waveguides and the junction waveguides should also be taken into account as, for instance, the latter waveguides will limit the number of *arms* that one can connect at the center. Initial numerical studies were carried using the structure from Fig. 1b for 3, 4 and 5 port configurations. From this, an increase in the number of ports caused a red shifted perfect splitting frequency (i.e., a shift towards the cut-off frequency of the outer waveguides $f_{c_w}$, not shown). Based on this, to be able to increase the number of *arms* (and hence ports) connected at the junction and to prevent the perfect splitting frequency from falling below $f_{c_w}$ (which would prevent perfect splitting from occurring as all outer and junction waveguides would be operating below their cut-off frequency of their fundamental mode), in this subsection both the outer and junction waveguide cut-off frequencies are modified to $f_{c_w} = 1.0$ GHz and $f_{c_x} = 1.8$ GHz, respectively. As observed, $f_{c_x}$ has now been increased compared to the structure discussed in Fig. 1-3, this is to enable the red shifting of the perfect splitting frequency when increasing *N* to be $> f_{c_w}$ and $< f_{c_x}$, as mentioned before, without exciting higher order modes in the outer waveguide (i.e., $f_{c_x} < 2f_{c_w}$ in our case).

Next, we can consider the length of the junction waveguides as a key parameter to be able to physically fit an increased number of *arms* connected to the junction. As the frequency at which perfect splitting occurs will be $< f_{c_x}$, it is the evanescent field inside the junction waveguides (working below their cut-off frequency $f_{c_x}$) that enables perfect splitting. Extending the length $\ell_x$ will reduce the evanescent field that reach the center of the junction waveguides, meaning that this length should be carefully chosen such that evanescent coupling to the junction still exist. A detailed numerical study of the influence of the length and height of the cross-section of the hollow for the junction waveguides (dimensions $\ell_x$ and $b_x = a_x \Delta h$, respectively) is presented in the supplementary materials. From those results, increasing $\ell_x$ cause the response of the structure to be more narrowband, a response that can be overcome by, for instance, increasing $\Delta h$. Interestingly, this means that narrow hollow waveguides are not required in our case to enable perfect splitting of EM signals. From now on, to evaluate perfect splitting response using $N$ port structures and to apply them for computing processes in the next section, the dimensions $\ell_x = 3\lambda_{c_x}$ and $\Delta h = 0.5$ are used.

The numerical results of the out-of-plane $E_y$- field distribution using the proposed interconnected rectangular waveguides for $N$ = 3, 5 and 6 *arms*, are shown in the first column of Fig 4. These results are calculated at the frequency where perfect splitting occurs for each value of $N$, respectively. As observed, the field distributions resemble the one shown for $N$ = 4 in Fig. 1, as expected. Scattering parameters for each of the $N$-ports are presented in the centre column of Fig. 4 with additional zoomed plots around the perfect splitting resonance in the right column of the same figure. As expected, the perfect splitting frequency is red shifted when increasing $N$, as discussed above. Moreover, the splitting ratios ($|S_{11}|$:$|S_{N1}|$) for the structures with $N$ = 3, 5, and 6 are ~1:2, ~3:2, and ~2:1 respectively (see exact values in the caption from Fig. 4). These results are in agreement with the theoretical values from Eq. 1, demonstrating how perfect splitting can be achieved using the proposed structure for $N$ ports. As shown in Fig. 4, here we have considered $N$ = 6 as the maximum value. This could be increased by, for instance, carefully increasing the length of the junction waveguides in order to fit more *arms* while making sure that the coupled evanescent field still reaches the junction. However, increasing $N$ will red shift even more the perfect splitting resonance, meaning that the dimensions of the outer and junction waveguide cross-sections may need to be modified (this is out of the scope of the present work). The proposed structure for perfect splitting of EM signals is exploited in the next section for fundamental computing processes including routing of information.

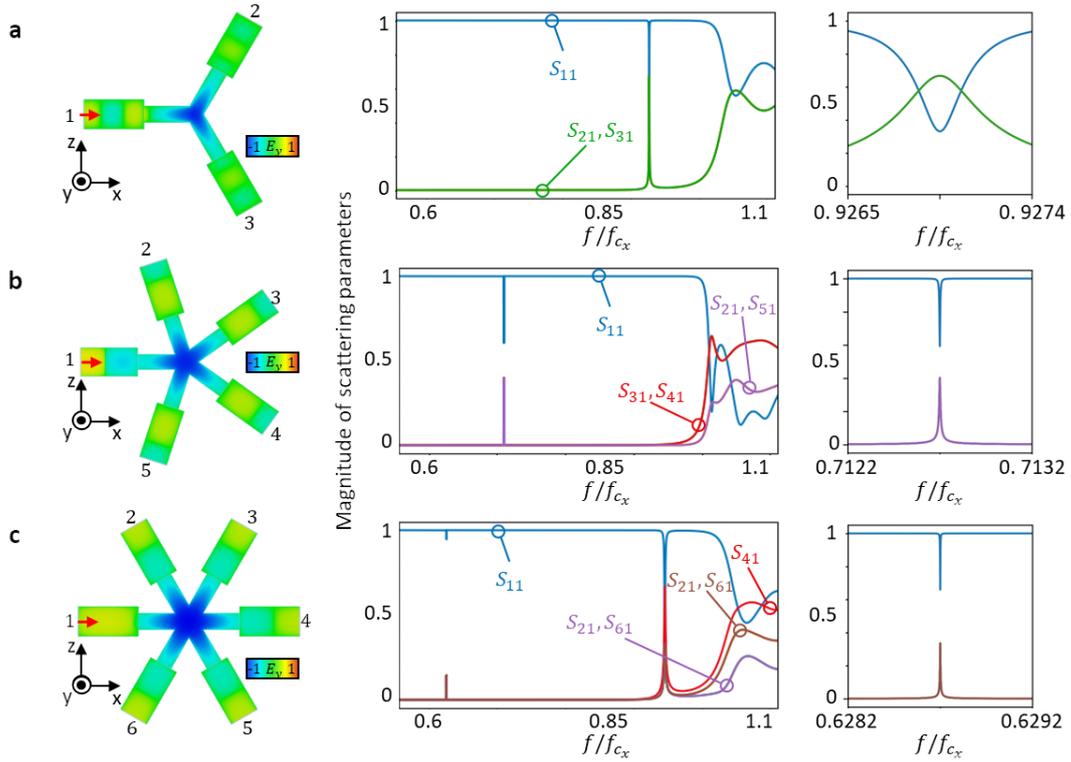

**Fig. 4| Perfect splitting with $N$-port interconnected rectangular waveguides.** (a-c) (left column) $E_y$-field distribution on the $xz$ plane ($H$-plane) for the $N = 3$, 5, 6-port structures, respectively. (Middle column), scattering parameters for the full spectral window and for a (right column) narrow frequency range around each perfect splitting resonance from the middle column. The values of reflection (towards port 1) and transmission (towards each of the output ports, ports 2 to $N$) coefficients for the $N = 3, 5, 6$-port structures are ($|\rho| = 0.3313$, $|\gamma| = 0.6669$), ($|\rho| = 0.5929$, $|\gamma| = 0.4025$), and ($|\rho| = 0.6599$, $|\gamma| = 0.3357$) respectively. These results are in good agreement with the expected values of perfect splitting in ideal TLs as discussed in section IIA using $|\rho| = |(2 - N)N^{-1}|$ and $|\gamma| = |2N^{-1}|$.

## IV. Exploiting perfect splitting for computing: examples

In the previous sections, perfect splitting of EM signals was achieved using interconnected rectangular waveguides. Here, it is shown how this performance can be used in fundamental computing operations such as routing of information and linear comparators. We have recently shown how such operations can be performed using arrays of parallel plate waveguides as ideal TLs [30], [31]. Here, we demonstrate that it is also possible to exploit rectangular waveguides, which are technologically important devices. Interestingly, while here we focus on two fundamental computing processes, perfect splitting of EM signals has shown to be an interesting mechanism to design, for instance, photonic structures to calculate temporal derivatives or the solution of partial differential equations [18], [34], [47], [48]. We envision that our proposed structures from the previous sections could also be used in this realm. A study that is currently under development and will be presented elsewhere.

### A. Routing of EM signals

For this scenario, a 4-port ($N = 4$) configuration of interconnected rectangular waveguides is used (as discussed in Fig. 1b). The dimensions of the outer and junction waveguides are the same as those shown in Fig. 4 (for $f_{c_w} = 1.0$ GHz and $f_{c_x} = 1.8$ GHz). To enable routing of information from one port to another port, we make use of the linear interaction between two incident signals applied from two different ports. Different scenarios considering input signals from ports 1,2 and ports 1,3 are shown in Fig. 5a,b and Fig. 5b,c, respectively, using different phases between the two incident signals.

Let us first focus on the case shown in Fig. 5a where two incident signals are applied (one from port 1, and one from port 2). These two signals are in-phase, as represented in Fig. 5a(i) as solid red and blue waves, respectively. When working at the perfect splitting resonance, as described in the previous sections, each incident signal will split into four signals (8 in total). Each of the new generated signals are depicted as dashed lines in the same Fig. 5a(i). All these scattered waves will have the same magnitude, with a phase difference between the transmitted and reflected signals of $\pi$. Then, the final output will simply be defined by Eq. 1, i.e., the signals traveling towards ports 1-4 after both incident signals have passed the junction will be the linear superposition of all the scattered signals. As a result, some signals will constructively/destructively interfere due to them being in-/out-of-phase. This is schematically shown Fig. 5a(i). As observed, no output signals are obtained traveling towards ports 1 and 2, while clear output signals exist traveling towards port 3 and 4. This behavior is also demonstrated in Fig. 5a(ii) where the numerical results of the $E_y$-field distribution is presented at the perfect splitting resonant frequency. Finally, the combined scattering parameters when using two input signals (i.e., the $F$-parameters, representing the output spectra at a port when considering the two incident signals as a simultaneous excitation [49], [50]) is presented in Fig. 5a(iii) where it can be seen how full transmission (value of ~1) is obtained at ports 3 and 4 when working at the first resonance ($f/f_{c_x}$~0.8125). To verify these results, one can apply Eq. 1 with an input vector of signals $x = [1, 1, 0, 0]$, resulting in the following output vector of signals $y$:

$$y = \begin{bmatrix} -0.5 & 0.5 & 0.5 & 0.5 \\ 0.5 & -0.5 & 0.5 & 0.5 \\ 0.5 & 0.5 & -0.5 & 0.5 \\ 0.5 & 0.5 & 0.5 & -0.5 \end{bmatrix} \begin{bmatrix} 1 \\ 1 \\ 0 \\ 0 \end{bmatrix} = \begin{bmatrix} 0 \\ 0 \\ 1 \\ 1 \end{bmatrix} \quad (10)$$

which is in agreement with the values from Fig. 5a(iii). Conversely, consider now two input signals applied from ports 1 and 2 with a phase difference of $\pi$, $x = [1, -1, 0, 0]$. This scenario is schematically shown in Fig. 5b. Following the same process, as in Fig. 5a(i), the final transmitted signal towards each of the ports will be the linear superposition of all the scattered signals produced by each of the applied inputs. In so doing, the two scattered signals towards ports 3 and 4 are out-of-phase while those towards ports 1 and 2 are in-phase. As a result, output signals are only observed in port 1 and 2. This is again in line with Eq. 1 which predicts an output vector of signals $y = [-1, 1, 0, 0]^T$. The numerical results of this case are shown in Fig. 5b(ii,iii) for the $E_y$- field distribution at the perfect splitting frequency and the full spectral response of the *F* parameters, respectively. Further examples are shown in Fig. 5c,b using input signals applied from ports 1 and 3 being in-phase or out-of-phase, respectively. For these two cases, in-phase (Fig. 5c) $x = [1, 0, 1, 0]$ and out-of-phase (Fig. 5d) input signals, $x = [1, 0, -1, 0]$, the output vector of signals is $y = [0, 1, 0, 1]^T$ and $y = [-1, 0, 1, 0]^T$, respectively, after using Eq. 1. These results are again in agreement with numerical calculations, as shown in Fig. 5c,d(ii,iii), respectively.

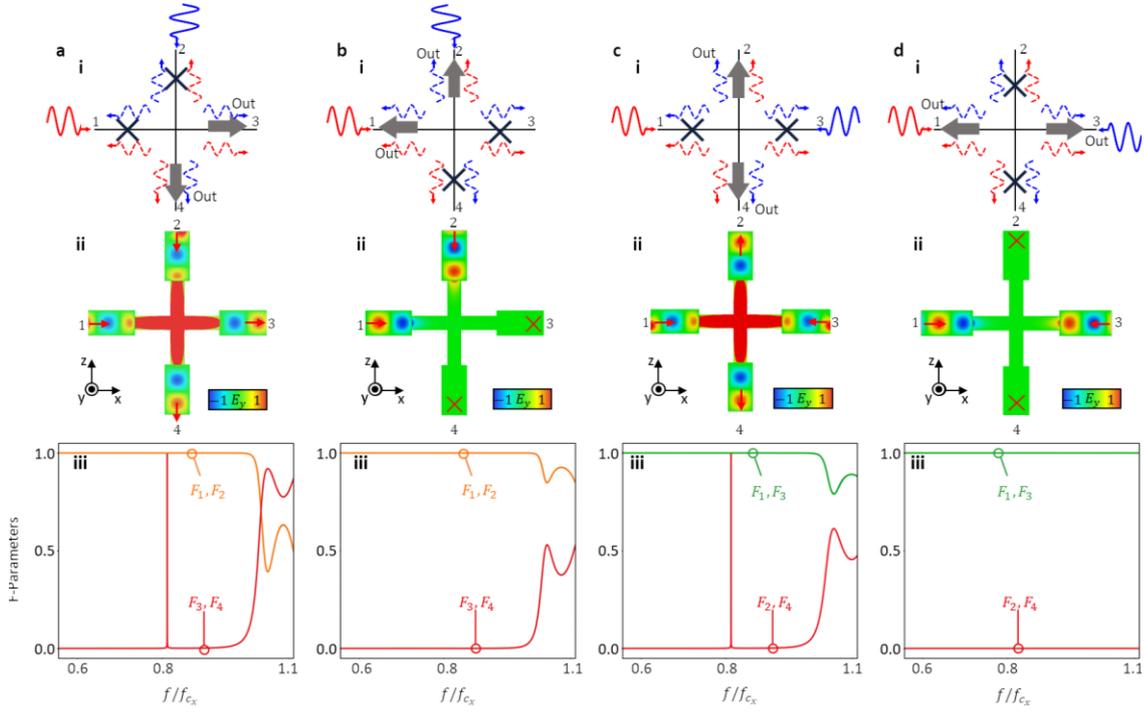

**Fig. 5| Demonstration of routing of information using a 4-port interconnected rectangular waveguide structure.** (a-b) Input signals from ports 1 and 2, (c-d) Input signals at ports 1 and 3, each for the in-phase and out-of-phase condition respectively. i) Schematic representation of the resulting interference of all the generated scattered signals generated after the input signals are perfectly split when passing the junction. ii) $E_y$-field distribution on the *xz* plane (*H*-plane). iii) *F*-parameters for the full spectral window, for each respective input and phase combination.

## B. Linear comparator

An interesting application of perfect splitting is that it can be used to perform the comparison between the amplitudes of two inputs [31]. Here, we demonstrate that such computing operation can also be achieved using the proposed perfect splitters via interconnected rectangular waveguides. This amplitude comparison operation is performed by utilizing the phase interaction in a 3-port perfect splitting structure. Consider an $N = 3$ interconnected rectangular waveguide junction where ports 1 and 2 are used as inputs. The amplitude and phase of the incident signals at each port are $(a_1^{i/o}, \phi_1^{i/o})$, $(a_2^{i/o}, \phi_2^{i/o})$, and $(a_3^{i/o}, \phi_3^{i/o})$, respectively, with superscripts $i/o$ representing input/output, respectively. Port 1 and 2 are used as inputs while Port 3 is considered an output port. Now, if the signals from ports 1 and 2 are simultaneously applied to the structure with $a_1^i = a_1^i$ but different phase such as $\phi_1^i - \phi_2^i = \pi$, and following Eq. 1, port 1 and 2 will see a reflected signal opposite phase (or polarity) while port 3 will see no signal. Such performance can then be exploited to compare the amplitude of two incident signals. As shown in [31] using ideal parallel plate waveguides, a comparator operation can be done by considering two square pulses with opposite polarities applied to port 1 and 2 in a $N = 3$ network of ideal waveguides. Then, one can look at the phase (polarity) of the output pulse towards port 3 and determine which input pulse had the largest amplitude. A demonstration of this scenario using our approach is shown in Fig. 6. First, we can consider the two incident signals defined by the input vector $x = [-1, 0.7, 0]$ (noting that both incident signals have different amplitude and opposite phase). Following Eq. 1, the output vector of signals is as follows:

$$y = \begin{bmatrix} -1/3 & 2/3 & 2/3 \\ 2/3 & -1/3 & 2/3 \\ 2/3 & 2/3 & -1/3 \end{bmatrix} \begin{bmatrix} -1 \\ 0.7 \\ 0 \end{bmatrix} = \begin{bmatrix} 0.8 \\ -0.9 \\ -0.2 \end{bmatrix} \quad (11)$$

which demonstrates that the phase of the output signal at port 3 is dominated by the phase of the input signal at port 1 (the larger amplitude input) since both values are negative. These results are in agreement with Fig. 6a(i) where the $E_y$- field distribution on the $xz$ plane is plotted. As observed, the phase of the output signal at port 3 is the same as output signal of port 2 but opposite to the output signal at port 1, in agreement with Eq. 11.

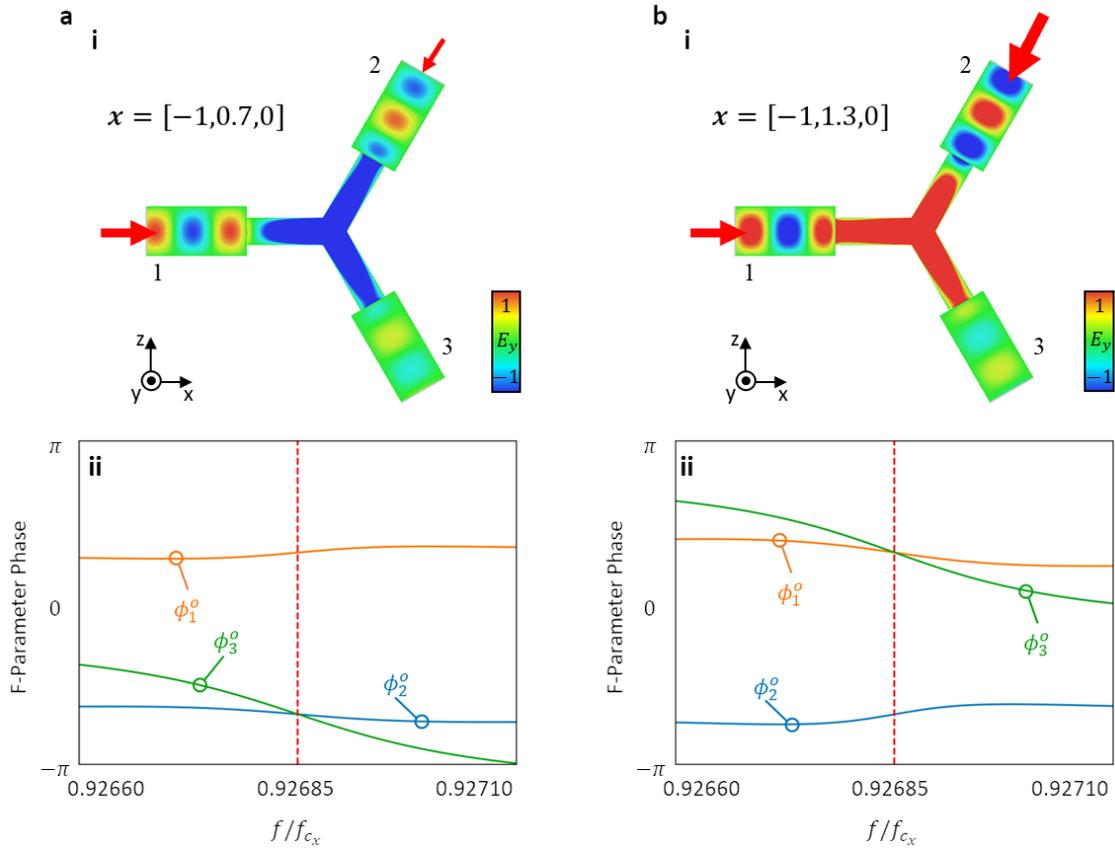

**Fig. 6| Demonstration of a $N=3$ interconnected rectangular waveguide junction amplitude comparator.** (a) Case 1, $a_1^i > a_2^i$. (b) Case 2, $a_1^i < a_2^i$. i) $E_y$-field distribution on the $xz$ plane (*H*-plane) at the resonant frequency shown as vertical dashed line in ii. ii) *F*-parameters for a narrow spectral window around the first resonance where perfect splitting occurs ($f/f_{c_x} \sim 0.9269$).

For completeness, the output phase spectra are shown in Fig. 6a(ii) where it is also seen how the phases are in agreement with Eq. 11. Finally, the opposite condition is also demonstrated in Fig. 6b where now the amplitude of the incident signal from port 2 is larger than that from port 1. Here, $x = [-1,1.3,0]$ resulting with an output vector $y = [1.2, -1.1, 0.2]^T$ from Eq. 1. Note that now the phase (polarity) of the signal at port 3 is the same as the phase of the input signal from port 2 which is the opposite that happened in the example from Fig. 6a, corroborating that now the incident signal from port 2 has higher amplitude than the incident one from port 1.

The corresponding numerical results for the $E_y$- field distribution at the resonant frequency along with the $F$-parameter are shown in Fig. 6b(i,ii), respectively, demonstrating a good agreement with the theoretical values.

**V. Conclusion**

In this manuscript, a structure to achieve perfect splitting of EM waves has been proposed and evaluated both theoretically and numerically. The proposed structure consisted of $N$ interconnected rectangular waveguides. It was demonstrated that perfect splitting cannot occur when considering just the junction of waveguides given that their width (in the direction perpendicular to the $E$-field) is in the order of the operational wavelength. To tackle this, additional outer waveguides, with a cut-off frequency below that of the junction waveguides, were considered. Thus, it was shown that it is possible to achieve perfect splitting of EM waves when working below the cut-off frequency of the junction waveguides, i.e. there is an evanescent field coupled to the junction waveguides, exciting a junction resonance. The proposed structure was then evaluated theoretically using TL models, demonstrating how the evanescent coupling and junction resonant frequency can be approximately modelled using parallel admittances in the equivalent circuit. Different scenarios of perfect splitting were demonstrated, including $N$ = 3, 4, 5 and 6 interconnected waveguides, showing excellent agreement with the theory of perfect splitters. Finally, the proposed structure was implemented to perform fundamental computing operations such as routing of information and comparison operations. We envision that the proposed mechanism for perfect splitting of EM waves could be implemented in large networks to perform, for instance, the solution of partial differential equations of derivatives both in the time and frequency domain, opening up further opportunities for computing with waves.

**Acknowledgements**

V.P.-P. and A. Y. would like to thank the support of the Leverhulme Trust under the Leverhulme Trust Research Project Grant scheme (No. RPG-2020-316). For the purpose of Open Access, the authors have applied a CC BY public copyright license to any Author Accepted Manuscript (AAM) version arising from this submission.

**Conflicts of interests**

The authors declare no conflicts of interests.

**Data availability**

The datasets generated and analysed during the current study are available from the corresponding author upon reasonable request.